\documentclass[allclo, onecollarge, square, comma, sort&compress, natbib]{svjour2}
\bibpunct{[}{]}{;}{n}{}{,} 
\smartqed  

\usepackage{mathrsfs}
\usepackage{amsmath}
\usepackage{amssymb}
\usepackage{bbold}
\usepackage{color}
\usepackage{graphicx}
\usepackage{soul}

\definecolor{darkgreen}{rgb}{0,0.5,0}
\definecolor{purple}{rgb}{0.5,0,0.5}
\definecolor{blue}{rgb}{0.0,0.0,0.50}
\definecolor{scarlet}{rgb}{1.0,0.2,0}

\usepackage[mathscr,scaled=1.15]{urwchancal}
\DeclareFontFamily{OT1}{pzc}{}
\DeclareFontShape{OT1}{pzc}{m}{it}%
{<-> s * [1.15] pzcmi7t}{}
\DeclareMathAlphabet{\mathpzc}{OT1}{pzc}{m}{it}

\journalname{Few-Body Systems}
\begin{document}

\title{Perspective on the origin of hadron masses}
\titlerunning{Perspective on the origin of hadron masses}        

\author{Craig D.\ Roberts}

\institute{Craig D.\ Roberts \at
            Physics Division, Argonne National Laboratory, Argonne, Illinois 60439, USA.
              \email{cdroberts@anl.gov}}

\date{Received: 10 June 2016 
}

\maketitle

\begin{abstract}
The energy-momentum tensor in chiral QCD, $T_{\mu\nu}$, exhibits an anomaly, \emph{viz}.\ $\Theta_0 :=T_{\mu\mu} \neq 0$.  Measured in the proton, this anomaly yields $m_p^2$, where $m_p$ is the proton's mass; but, at the same time, when computed in the pion, the answer is $m_\pi^2=0$.  Any attempt to understand the origin and nature of mass, and identify observable expressions thereof, must explain and unify these two apparently contradictory results, which are fundamental to the nature of our Universe.
Given the importance of Poincar\'e-invariance in modern physics, the utility of a frame-dependent approach to this problem seems limited.  That is especially true of any approach tied to a rest-frame decomposition of $T_{\mu\nu}$ because a massless particle does not possess a rest-frame.
On the other hand, the dynamical chiral symmetry breaking (DCSB) paradigm, connected with a Poincar\'e-covariant treatment of the continuum bound-state problem, provides a straightforward, simultaneous explanation of both these identities, and also a diverse array of predictions, testable at existing and proposed facilities.  From this perspective, $\langle \pi| \Theta_0 |\pi \rangle =0$ owing to exact, symmetry-driven cancellations which occur between one-body dressing effects and two-body-irreducible binding interactions in any well-defined computation of the forward scattering amplitude that defines this expectation value in the pseudoscalar meson.  The cancellation is incomplete in any other hadronic bound state, with a remainder whose scale is set by the size of one-body dressing effects.
\end{abstract}

\section{Introduction}
Classical chromodynamics (CCD) is a non-Abelian local gauge field theory.  As with all such theories formulated in four spacetime dimensions, no mass-scale exists in the absence of Lagrangian masses for the fermions.  There is no dynamics in a scale-invariant theory, only kinematics: the theory looks the same at all length-scales and hence there can be no clumps of anything.  Bound-states are therefore impossible and, accordingly, our Universe cannot exist.  Spontaneous symmetry breaking, as realised via the Higgs mechanism, does not solve this problem because normal matter is constituted from light-quarks, $u$ and $d$, and the masses of the neutron and proton, the kernels of all visible matter, are roughly 100-times larger than anything the Higgs can produce in connection with $u$- and $d$-quarks.  Consequently, the question of how did the Universe come into being is inseparable from the questions of how does a mass-scale appear and why does it have the value we observe?

%
Modern quantum field theories are not built simply on Lorentz invariance.  The effect of space-time translations must also be considered and thus enters the group of Poincar\'e transformations.  In this connection, consider the energy-momentum tensor in CCD, $T_{\mu\nu}$, which can always be made symmetric \cite{BELINFANTE1940449}.  Conservation of energy and momentum in a quantum field theory is a consequence of spacetime translational invariance, one of the family of Poincar\'e transformations.  Consequently,
\begin{equation}
\label{ConsEP}
\partial_\mu T_{\mu\nu} = 0 \,.
\end{equation}

Consider now a global scale transformation in the Lagrangian of the classical theory:
\begin{subequations}
\label{scaleT}
\begin{align}
\label{xxprime}
x & \to x^\prime = {\rm e}^{-\sigma}x\,, \\
A_\mu^a(x) & \to A_\mu^{a\prime}(x^\prime) = {\rm e}^{-\sigma} A_\mu^a({\rm e}^{-\sigma}x ) \,,\\
q(x) & \to q^\prime(x^\prime) = {\rm e}^{- (3/2) \sigma} q({\rm e}^{-\sigma}x )\,,
\end{align}
\end{subequations}
where $A_\mu^a(x)$, $q(x)$ are the gluon and quark fields.  The Noether current associated with this transformation is the dilation current
\begin{equation}
{\cal D}_\mu = T_{\mu\nu} x_\nu\,.
\end{equation}
In the absence of fermion masses, the classical action is invariant under Eqs.\,\eqref{scaleT}, \emph{i.e}.\ the theory is scale invariant, and hence
\begin{equation}
\partial_\mu {\cal D}_\mu  = 0  = [\partial_\mu T_{\mu\nu} ] x_\nu + T_{\mu\nu} \delta_{\mu\nu}  = T_{\mu\mu}\,,
\label{SIcQCD}
\end{equation}
where the last equality follows from Eq.\,\eqref{ConsEP}.  Plainly, the energy-momentum tensor is traceless in a scale invariant theory.\footnote{It is possible for a quantum field theory to be scale invariant but not conformally-invariant; but examples are rare.  In fact, there is no known example of a scale invariant but non-conformal field theory in four dimensions under a small set of seemingly reasonable assumptions.  For this reason, the terms are often used interchangeably in the context of quantum field theory, even though the scale symmetry group is smaller.  (See, e.g.\ Refs.\,\cite{Nakayama:2013is, Qualls:2015qjb}.)}
%
%

Massless CCD is not a meaningful framework in modern physics for many reasons; amongst them the fact that strong interactions in the Standard Model are empirically known to be characterised by a large mass-scale.  In quantising the theory, regularisation and renormalisation of (ultraviolet) divergences introduces a mass-scale.  This is ``dimensional transmutation'': mass-dimensionless quantities become dependent on a mass-scale, and this entails the violation of Eq.\,\eqref{SIcQCD}, \emph{i.e}. the appearance of the chiral-limit ``trace anomaly'':
\begin{equation}
\label{SIQCD}
T_{\mu\mu} = \beta(\alpha(\zeta))  \tfrac{1}{4} G^{a}_{\mu\nu}G^{a}_{\mu\nu} =: \Theta_0 \,,
\end{equation}
where $\beta(\alpha)$ is the QCD $\beta$-function, $\zeta$ is the renormalisation scale, $G^{a}_{\mu\nu}$ is the gluon field-strength tensor, and this expression assumes the  chiral limit for all current-quarks.

There is a simple, nonperturbative derivation of this identity, which eliminates any need for a diagrammatic or perturbative analysis \cite{gonsalvesphy522}.  Namely, under the scale transformation in Eq.\,\eqref{xxprime}, the mass-scale $\zeta \to {\rm e}^\sigma \zeta$.  Considering infinitesimal transformations of this type, it is straightforward to show:
\begin{equation}
\alpha \to \sigma\,  \alpha  \beta(\alpha)\,,\;
{\cal L} \to \sigma \, \alpha \beta(\alpha) \, \frac{\delta {\cal L}}{\delta \alpha}
\Rightarrow \partial_\mu {\cal D}_\mu = \frac{\delta {\cal L}}{\delta \sigma} = \alpha \beta(\alpha) \, \frac{\delta {\cal L}}{\delta \alpha}\,.
\end{equation}
In order to compute the final product, one may first absorb the gauge coupling into the gluon field, \emph{i.e}.\ express the action in terms of $\tilde A_\mu^{a} = g A_\mu^a$, in which case the running coupling appears only as an inverse multiplicative factor connected with the pure-gauge term:
\begin{equation}
 {\cal L}(\alpha)  = - \frac{1}{4\pi\alpha}\frac{1}{4} \tilde G_{\mu\nu}^a \tilde G_{\mu\nu}^a + \alpha\mbox{-independent~terms,}
\end{equation}
where $\tilde G_{\mu\nu}^a$ is the field-strength tensor expressed using $\tilde A_\mu^{a}$.  It then follows that
 \begin{equation}
T_{\mu\mu} = \partial_\mu {\cal D}_\mu = \alpha \beta(\alpha) \frac{\delta {\cal L}}{\delta \alpha} =  \alpha \beta(\alpha)\frac{1}{4\pi\alpha^2} \frac{1}{4} \tilde G_{\mu\nu}^a \tilde G_{\mu\nu}^a = \beta(\alpha)  \tfrac{1}{4} G^{a}_{\mu\nu}G^{a}_{\mu\nu}\,,
\end{equation}
\emph{viz}.\ Eq.\,\eqref{SIQCD} is recovered.

It is worth emphasising here that the appearance of a trace anomaly has nothing to do with the non-Abelian nature of QCD.  Indeed, it will be apparent from the derivation just sketched that quantum electrodynamics (QED) must also possess a trace anomaly.  However, QED is nonperturbatively undefined: four-fermion operators become relevant in strong-coupling QED and must be included in order to obtain a well-defined (albeit trivial) continuum limit (see, \emph{e.g}.\, Refs.\,\cite{Rakow:1990jv, Reenders:1999bg, Akram:2012jq}).  As a consequence, QED does not have a chiral limit.  The QED trace anomaly is only meaningful in perturbation theory and its scale is determined by the Higgs mechanism.
%

In the presence of nonzero current-quark masses, Eq.\,\eqref{SIQCD} becomes
\begin{equation}
T_{\mu\mu} = \tfrac{1}{4} \beta(\alpha(\zeta)) G^{a}_{\mu\nu}G^{a}_{\mu\nu}
+ [1+\gamma(\alpha(\zeta))]\sum_f m_f^\zeta \, \bar q_f q_f\,,
\end{equation}
where $m_f^\zeta$ are the current-quark masses and $[1+\gamma(\alpha)]$ is the analogue for the dressed-quark running-mass of $\beta(\alpha)$ for the running coupling.  (In fact, $\gamma(\alpha)$ is the anomalous dimension of the current-quark mass in QCD.)
It is notable that in the massive-case the trace anomaly is not homogeneous in the running coupling, $\alpha(\zeta)$.  Consequently, renormalisation-group-invariance does not entail form invariance of the right-hand-side (rhs) \cite{Tarrach:1981bi}.  This is important because discussions typically assume (perhaps implicitly) that all operators and identities are expressed in a partonic basis, \emph{viz}.\ using simple field operators that can be renormalised perturbatively, in which case the hadronic state-vector represents an extremely complicated wave function.  That perspective is not valid at renormalisation scales $\zeta \lesssim m_p$, where $m_p$ is the proton mass; and this is where a metamorphosis from parton-basis to quasiparticle-basis may occur: under reductions in resolving scale, $\zeta$, light partons evolve into heavy dressed-partons, corresponding to complex superpositions of partonic operators; and using these dressed-parton operators, the wave functions can be expressed in a relatively simple form.  (A relevant illustration is discussed in Refs.\,\cite{Finger:1981gm, Adler:1984ri}.)

\section{Magnitude of the Scale Anomaly: Mass and Masslessness}
\subsection{A binary problem}
Simply knowing that a trace anomaly exists does not deliver a great deal: it only indicates that there is a mass-scale.  The crucial issue is whether or not one can compute and/or understand the magnitude of that scale.

One can certainly measure the size of the scale anomaly, for consider the expectation value of the energy-momentum tensor in the proton:
\begin{equation}
\label{EPTproton}
\langle p(P) | T_{\mu\nu} | p(P) \rangle = - P_\mu P_\nu\,,
\end{equation}
where the rhs follows from the equations-of-motion for an asymptotic one-particle proton state.  At this point it is clear that, in the chiral limit, \begin{subequations}
\label{anomalyproton}
\begin{align}
\langle p(P) | T_{\mu\mu} | p(P) \rangle  = - P^2 & = m_p^2\\
& = \langle p(P) |  \Theta_0 | p(P) \rangle\,;
\end{align}
\end{subequations}
namely, there is a clear sense in which it is possible to say that the entirety of the proton mass is produced by gluons.  The trace anomaly is measurably large; and that property must logically owe to gluon self-interactions, which are also responsible for asymptotic freedom.

This is a valid conclusion.  After all, what else could be responsible for a mass-scale in QCD?  QCD is all about gluon self-interactions; and it's gluon self-interactions that (potentially) enable one to rigorously (nonperturbatively) define the expectation value in Eq.\,\eqref{anomalyproton}.  On the other hand, it's only a sensible conclusion when the operator and the wave function are defined at a resolving-scale $\zeta \gg m_p$.  It will be necessary to return to this point.

There is also another issue, which can be exposed by returning to Eq.\,\eqref{EPTproton} and replacing the proton by the pion:
\begin{equation}
\label{EPTpion}
\langle \pi(q) | T_{\mu\nu} | \pi(q) \rangle = - q_\mu q_\nu\,.
\end{equation}
Then, in the chiral limit:
\begin{align}
\label{anomalypion}
& \langle \pi(q) |  \Theta_0 | \pi (q) \rangle = 0
\end{align}
because the pion is a massless Nambu-Goldstone mode.  Equation\,\eqref{EPTpion} could mean that the scale anomaly vanishes trivially in the pion state, \emph{viz}.\ that gluons and their self-interactions have no impact within a pion because each term in the expression of the operator vanishes when evaluated in the pion. However, that is a difficult way to achieve Eq.\,\eqref{anomalypion}.  It is easier, perhaps, to imagine that Eq.\,\eqref{anomalypion} owes to cancellations between different operator-component contributions.  Of course, such precise cancellation should not be an accident.  It could only arise naturally because of some symmetry and/or symmetry-breaking pattern; and, as will be argued below, that is precisely the manner by which Eq.\,\eqref{anomalypion} is realised.

Equations\,\eqref{anomalyproton} and \eqref{anomalypion} present a quandary, which highlights that no understanding of the origin of the proton's mass can be complete unless it simultaneously explains the meaning of Eq.\,\eqref{anomalypion}.  Given that a massless particle doesn't have a rest-frame, any approach based on a rest-frame decomposition of the energy-momentum tensor (\emph{e.g}.\ Refs.\,\cite{Ji:1994av, Ji:1995sv}) cannot readily be useful in this dual connection.

\subsection{Frame independent resolution}
Both Eqs.\,\eqref{anomalyproton} and \eqref{anomalypion} are Poincar\'e invariant statements.  Quantum field theories are the only known realisation of the Poincar\'e algebra in quantum mechanics with a particle interpretation.  This entails that asymptotic one-particle states are characterised by just two invariants \cite{Coester:1992cg}, \emph{i.e}.\ the eigenvalues of $M^2$ and $W^2$, where the former is the mass-squared operator and $W_\mu$ is the Pauli-Lubanski four-vector.  The eigenvalues of the mass-squared operator, $m^2$, need no further explanation.

Concerning the Pauli-Lubanski four-vector, note that, by definition, $W_\mu$ contains no information about orbital angular momentum: it is sensitive only to the total-spin of the system.  For massive particles, the eigenvalues of $W^2$ are the products $m^2 j(j+1)$, where $j(j+1)$ is the rest-frame eigenvalue of $\vec{J}\cdot\vec{J}$, with $\vec{J}$ being the total angular momentum operator, \emph{viz}.\ $j$ is the particle's spin.  Since $W^2$ is Poincar\'e invariant, then the eigenvalues of $\vec{J}\cdot\vec{J}$ are the same in any frame.

In the massless case, the eigenvalues of $W^2$ vanish, the Pauli-Lubanski four-vector is proportional to the four-momentum, and the constant of proportionality is the particle's helicity.  Consequently, massless one-particle states are labelled by their helicity which, for fermions, is related to their chirality.  Massless fermions are either left-handed or right-handed and no Poincar\'e transformation can alter the assignment.

These statements entail that the only unambiguous labels that can be attached to a hadron state are its mass and spin\,$:=\,$total-angular-momentum; and different observers, characterised by distinct reference frames, will only necessarily agree on these two quantities.  Hence, \emph{e.g}.\ no two observers need necessarily agree on a massive hadron's polarisation.
Furthermore, in a quantum field theory, no separation of the total angular momentum into a sum $L+S$ can be Poincar\'e invariant.  Such a separation is frame-dependent, and that includes frames made distinct by boosts.  Consequently, no two observers (or calculators) need agree on the values of $\langle L \rangle$ and $\langle S\rangle$, even though they will agree on $\langle W^2 \sim (L+S)^2 \rangle$.  This fact lies at the heart of the so-called ``spin-crisis'', which could therefore have been avoided.
Finally, even using light-front quantisation, both the nature and size of contributions from constituents to any observable property of the composite hadron itself change with resolving scale, $\zeta$.
It seems, therefore, that a simultaneous elucidation of the meaning and consequences of Eqs.\,\eqref{anomalyproton} and \eqref{anomalypion} cannot satisfactorily be achieved by focusing on a particular reference frame and that the picture one finds most satisfactory will likely depend on the scale at which the resolution is presented.

It is also worth recalling that CCD is still a non-Abelian local gauge theory.  Consequently, the concept of local gauge invariance persists.  However, without a mass-scale there is no confinement.  For example, three quarks can be prepared in a colour-singlet combination and colour rotations will keep the three-body system colour neutral; but the quarks involved need not have any proximity to one another.  Indeed, proximity is meaningless because all lengths are equivalent in a scale invariant theory.  Hence, the question of ``Whence mass''? is equivalent to ``Whence a mass-scale?'', which is equivalent to ``Whence a confinement scale?''.  Thus, understanding the origin, Eq.\,\eqref{anomalyproton}, and absence, Eq.\,\eqref{anomalypion}, of mass in QCD is quite likely inseparable from the task of understanding confinement; and existence, alone, of a scale anomaly answers neither question.

\subsection{Value of a hadronic scale}
As noted above, the energy-momentum tensor is typically considered in connection with partonic operators, which are simple and can be computed perturbatively.  In this approach, however, the wave functions are extremely complicated; and they have never been computed in four dimensions, even approximately.\footnote{The problem of QCD in two-dimensions is discussed in Ref.\,\cite{Hornbostel:1988fb}.  However, two-dimensional theories have little in common with their four-dimensional counterparts, so the analysis in Ref.\,\cite{Hornbostel:1988fb} has not yet led to much progress in four-dimensional QCD.}  The partonic perspective is valid for $\zeta \gg \zeta_2 := 2\,$GeV.  It might be valid at smaller scales, too; but it cannot be used for $\zeta \lesssim \zeta_c \approx 0.5\,$GeV, which corresponds to the mass-scale at which coloured two-point functions in QCD exhibit an inflection point.  The value $\zeta_c$ is known from continuum- and lattice-QCD analyses (see, \emph{e.g}.\ Ref.\,\cite{Aguilar:2015bud} and citations therein).

Suppose one is working at renormalisation scales $\zeta \geq \zeta_2$.  If one begins to speak about a wave function with a quantum mechanical interpretation, then it is necessary to employ light-front quantisation.  In this case, a hadron wave function, $\Psi$, is independent of the hadron's four-momentum and, in principle, has a meaningful Fock-space decomposition \cite{Brodsky:1997de}.  On the other hand, it is not independent of the renormalisation scale: $\Psi = \Psi(\zeta)$, such that the relative strength of each Fock-space element changes with $\zeta$.  These changes are described by QCD evolution equations (\emph{e.g}.\ Refs.\,\cite{Gribov:1971zn, Gribov:1972, Dokshitzer:1977sg, Altarelli:1977, Lepage:1979zb, Efremov:1979qk, Lepage:1980fj}), which are known to some low/finite-order in perturbative QCD.  A method for evolving the light-front Hamiltonian to scales $\zeta \lesssim \zeta_c$ has long been sought, \emph{e.g}. Ref.\,\cite{Wilson:1994fk}.  In this case, as already noted, the operators become complicated, describing strongly-dressed quasi-particles; but the wave functions become simple, expressed in terms of a few quasi-particle degrees of freedom.  The light-front holographic approach to QCD \cite{Brodsky:2014yha} and Dyson-Schwinger equation (DSE) analyses \cite{Cloet:2013jya, Roberts:2015lja, Horn:2016rip} may be viewed as modern attempts to achieve the goal identified in Ref.\,\cite{Wilson:1994fk}.

It is evident now that the question: ``How does one understand a hadron's mass in terms of the contributions from its constituents and their interaction dynamics?'' is ill-posed because the meaning of the question depends on the energy-scale being used to probe the hadron and the reference frame to which the question is addressed.

The DSEs are useful here because they provide a symmetry-preserving (and hence Poincar\'e covariant) framework with a traceable connection to the Lagrangian of QCD.
The known limitation of this approach is the need to employ a truncation in order to define a tractable continuum bound-state problem.  That truncation might also involve an \emph{Ansatz} for the infrared behaviour of one or more coloured Schwinger functions, although that need is passing \cite{Binosi:2014aea}.
Concerning truncation, much has been learnt in the past twenty years, so that one may now separate DSE predictions into three classes:
(\emph{A}) model-independent statements about QCD;
(\emph{B}) illustrations of such statements using well-constrained model elements and possessing a traceable connection to QCD;
(\emph{C}) analyses that can fairly be described as QCD-based but whose elements have not been computed using a truncation that preserves a systematically-improvable connection with QCD.

\begin{figure}[t]
\centerline{\includegraphics[width=0.45\textwidth]{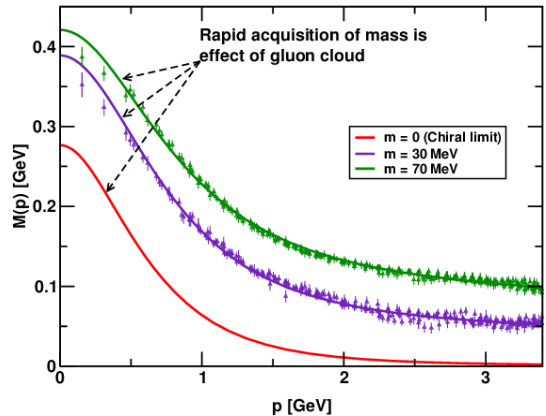}}
\caption{\label{gluoncloud} \small
Renormalisation-group-invariant dressed-quark mass function, $M(p)$: \emph{solid curves} -- DSE results, explained in Refs.\,\protect\cite{Bhagwat:2003vw,Bhagwat:2006tu}, ``data'' -- numerical simulations of lattice-regularised QCD \protect\cite{Bowman:2005vx}.  (\emph{N.B}.\ $m=70\,$MeV is the uppermost curve and current-quark mass decreases from top to bottom.)  The current-quark of perturbative QCD evolves into a constituent-quark as its momentum becomes smaller.  The constituent-quark mass arises from a cloud of low-momentum gluons attaching themselves to the current-quark.  This is dynamical chiral symmetry breaking (DCSB): an essentially nonperturbative effect that generates a quark \emph{mass} \emph{from nothing}; namely, it occurs even in the chiral limit.  The size of $M(0)$ is a measure of the magnitude of the QCD scale anomaly in $n=1$-point Schwinger functions.
}
\end{figure}

The dressed-quark mass-function depicted in Fig.\,\ref{gluoncloud} is a Class-A prediction combined with a Class-B illustration: owing to gluon self-interactions in QCD, massless quarks acquire a momentum-dependent mass-function, which is large in the infrared.  This figure is a clean demonstration of the scale anomaly at work: zero parton-mass becomes a mass-function, whose value depends on the scale at which the subsystem is probed.

\section{Confinement and Dynamical Chiral Symmetry Breaking}
\subsection{Millennium Prize}
The Clay Mathematics Institute has published a Millennium Prize Problem \cite{Jaffe:Clay}, whose solution will contain a proof of confinement in pure Yang-Mills theory.  The opening statement of the challenge reads: ``Prove that for any compact simple gauge group $G$, a non-trivial Yang-Mills theory exists on $\mathbb{R}^4$ and has a mass gap $\Delta >0$.''
There is strong evidence supporting the existence of a mass gap, found especially in the fact that numerical simulations of lattice-regularised QCD (lQCD) predict $\Delta \gtrsim 1.5\,$GeV \cite{McNeile:2008sr}.  However, even allowing for the Higgs mechanism, with this value of the mass gap, one computes $\Delta^2/m_\pi^2 \gtrsim 100$; and, therefore, a conundrum presents itself: can the mass-gap in pure Yang-Mills theory really play any role in understanding confinement when DCSB, driven by the same dynamics, ensures the existence of an almost-massless strongly-interacting excitation in our Universe?  If the answer is not \emph{no}, then it must at least be that one cannot claim to provide a pertinent understanding of confinement without simultaneously explaining its connection with DCSB.  The pion must play a critical role in any explanation of confinement in the Standard Model; and any discussion that omits reference to the pion's role is \emph{practically} \emph{irrelevant}.

From this perspective, the potential between infinitely-heavy quarks measured in simulations of quenched lQCD -- the so-called static potential \cite{Wilson:1974sk} -- is disconnected from the question of confinement in our Universe.  This is because light-particle creation and annihilation effects are essentially nonperturbative in QCD, so it is impossible in principle to compute a quantum mechanical potential between two light quarks \cite{Bali:2005fu, Prkacin:2005dc, Chang:2009ae}.  It follows that there is no flux tube in a Universe with light quarks and consequently that the flux tube is not the correct paradigm for confinement.\footnote{It is sometimes argued that hadron bound-states lie on linear Regge trajectories and there must therefore be a flux tube.  However, empirical evidence for the existence of such towers of states tied to parallel, linear trajectories is poor, \emph{e.g}.\ Refs.\,\cite{Tang:2000tb, Masjuan:2012gc}; the potential required to produce such trajectories is dependent both on the frame and the quantisation-scheme employed, \emph{i.e}.\ their appearance and nature is strongly dependent on the model used; and no approach whose parameters can rigorously be connected with real-world QCD has ever produced such trajectories.}

DCSB is the key here.  It ensures the existence of (pseudo-)Nambu-Goldstone modes; and in the presence of these modes, it is unlikely that any flux tube between a static colour source and sink can have a measurable existence.  To explain this statement, consider such a tube being stretched between a source and sink.  The potential energy accumulated within the tube may increase only until it reaches that required to produce a particle-antiparticle pair of the theory's pseudo-Nambu-Goldstone modes.  Simulations of lQCD show \cite{Bali:2005fu, Prkacin:2005dc} that the flux tube then disappears instantaneously along its entire length, leaving two isolated colour-singlet systems.  The length-scale associated with this effect in QCD is $r_{\not\sigma} \simeq (1/3)\,$fm and hence if any such string forms, it would dissolve well within a hadron's interior.

An alternative realisation associates confinement with dramatic, dynamically-driven changes in the analytic structure of QCD's coloured propagators and vertices.  That leads these coloured $n$-point functions to violate the axiom of reflection positivity and hence forces elimination of the associated excitations from the Hilbert space associated with asymptotic states \cite{GJ81}.  This is a sufficient condition for confinement \cite{Stingl:1985hx, Krein:1990sf, Hawes:1993ef, Roberts:1994dr, Cloet:2013jya}.  It should be noted, however, that the appearance of such alterations when analysing some truncation of a given theory does not mean that the theory itself is truly confining: unusual spectral properties can be introduced by approximations, leading to a truncated version of a theory which is confining even though the complete theory is not, \emph{e.g}.\ Refs.\,\cite{Krein:1993jb, Bracco:1993cy}.  Notwithstanding exceptions like these, a computed violation of reflection positivity by coloured functions in a veracious treatment of QCD does express confinement.  Moreover, via this mechanism, confinement is achieved as the result of an essentially dynamical process.

Figure~\ref{gluoncloud} highlights that quarks acquire a running mass distribution in QCD; and this is also true of gluons (see, \emph{e}.\emph{g}.\ Refs.\,\cite{Aguilar:2008xm, Aguilar:2009nf, Boucaud:2011ug, Pennington:2011xs, Ayala:2012pb, Binosi:2014aea, Aguilar:2015bud}).  The generation of these masses leads to the emergence of a length-scale $\varsigma \approx 0.5\,$fm, whose existence and magnitude is evident in all existing studies of dressed-gluon and -quark propagators, and which characterises the dramatic change in their analytic structure that has just been described.  In models based on such features \cite{Stingl:1994nk}, once a gluon or quark is produced, it begins to propagate in spacetime; but after each ``step'' of length $\varsigma$, on average, an interaction occurs so that the parton loses its identity, sharing it with others.  Finally a cloud of partons is produced, which coalesces into colour-singlet final states.  This picture of parton propagation, hadronisation and confinement can be tested in experiments at modern and planned facilities \cite{Accardi:2009qv, Dudek:2012vr, Accardi:2012qutS}.

\subsection{Nambu-Goldstone modes}
Returning to Eq.\,\eqref{anomalypion}, suppose now that the Higgs mechanism is active, in which case
\begin{equation}
\label{anomalypionm}
\langle \pi(q) | \Theta |\pi(q)\rangle
= (m_u^\zeta + m_d^\zeta) \frac{\rho_\pi^\zeta}{f_\pi}
\end{equation}
where \cite{Maris:1997hd, Brodsky:2012ku} $f_\pi$ is the pseudovector projection of pion's Bethe-Salpeter wave function onto the origin in configuration space (pion's leptonic decay constant) and $\rho_\pi^\zeta$ is the analogous pseudoscalar projection.\footnote{The extension to charge-neutral pseudoscalar mesons is readily accomplished following Ref.\,\cite{Bhagwat:2007ha}.}

Plainly,``heavy''-quarks play no role in generating $m_\pi$ via the trace anomaly, so they're probably not going to play a large part in $m_p$.  In fact, Class-B and -C DSE analyses indicate that, with physical values of the Higgs-generated current-masses, explicit mass terms for $u$- and $d$-quarks contribute $f_{u+d}=6$\% of the proton mass and that of the $s$-quark, just $f_s=2$\%  \cite{Flambaum:2005kc, Holl:2005st, Cloet:2008fw}.  These predictions are consistent with contemporary lQCD results \cite{Shanahan:2012wh, Bali:2016lvx}.  There is no concrete reason to expect a material contribution from quarks with larger current-masses and arguments to suggest that they are small.  For example, simple scaling within the frameworks used to produce the preceding fractions suggests that $c$-quarks contribute $\lesssim 0.1$\% of $m_p$, \emph{viz}.
\begin{equation}
f_c = f_s \, \frac{f_K^2}{f_D^2} \, \frac{m_K^2}{m_D^2} \, \frac{M_c}{M_s} \approx 0.1\,,
\end{equation}
where $f_{K,D}$ are meson leptonic decay constants, $m_{K,D}$ are meson masses, and $M_{s,c}$ are constituent-like quark masses \cite{Ivanov:1998ms, El-Bennich:2016bno}.

The rhs of Eq.\,\eqref{anomalypionm} only involves light-quark current-masses.  However, this does \emph{not} mean that gluons contribute nothing to the pion's mass because both $f_\pi$ and $\rho_\pi^\zeta$ are order parameters for DCSB, confinement leads to DCSB, gluons generate confinement, and consequently gluons are at least responsible for the strength of $\rho_\pi^\zeta/f_\pi$ and hence the rate at which $m_\pi$ increases with current-quark mass.  Gluons actually play a far greater role than this, as will be discussed further below.

In addressing the pion mass and its connection with DCSB, it is crucial to grasp some basic identities in QCD.  To that end, consider the pion's Bethe-Salpeter amplitude:
\begin{equation}
\Gamma_\pi(k;P) = i \gamma_5 \left[ E_\pi(k;P) + \gamma\cdot P \, F_\pi(k;P)   + \gamma\cdot k k\cdot P \, G_\pi(k;P) + \sigma_{\mu\nu} k_\mu P_\nu\, H_\pi(k;P) \right]\,. \label{Gammapi}
\end{equation}
In the chiral limit the axial-vector Ward-Green-Takahashi identity entails the following array of Goldberger-Treiman relations \cite{Maris:1997hd, Qin:2014vya}:
\begin{subequations}
\label{gGTrelations}
\begin{align}
f_\pi^0 E_{\pi}^0(y,w=0;P^2=0) &= B_0(y) \,, \label{BGTrelation}\\
F_A^0(y,w=0;P^2=0)
 + 2 f_\pi^0 F_\pi^0(y,w=0;P^2=0) &= A_0(y) \,, \\
  G_A^0(y,w=0;P^2=0)
+ 2 f_\pi^0 G_\pi^0(y,w=0;P^2=0) &= 2 A_0^\prime(y) \,, \\
 H_A^0(y,w=0;P^2=0)
+ 2 f_\pi^0 H_\pi^0(y,w=0;P^2=0) &= 0 \,,
\end{align}
\end{subequations}
where $y=k^2$, $w=k\cdot P$, $F_A^0$, $G_A^0$, $H_A^0$ are regular functions that appear in the axial-vector vertex, and the chiral-limit dressed-quark propagator is
\begin{equation}
S_0(p)
 = 1/[i \gamma\cdot p \, A_0(p^2,\zeta^2) + B_0(p^2,\zeta^2)] \label{SABform}
 = Z_0(p^2,\zeta^2)/[i \gamma\cdot p  + M_0(p^2)]\,.
\end{equation}
It follows that in the chiral limit, DCSB is a sufficient and necessary condition for the appearance of a massless pseudoscalar bound-state that dominates the axial-vector vertex on $P^2\simeq 0$ and whose constituents are described by a momentum-dependent mass-function, which may be arbitrarily large.  Furthermore, the appearance of a dynamically-generated nonzero mass-function in the solution of QCD's chiral-limit one-quark problem entails, through Eqs.\,\eqref{gGTrelations} in general, and Eq.\,\eqref{BGTrelation} in particular, that the isospin-nonzero pseudoscalar two-body problem is solved, well-nigh completely and without additional effort, once the solution to the one-body dressed-quark problem is known; and, moreover, that the quark-level Goldberger-Treiman relation in Eq.\,\eqref{BGTrelation} is the most basic expression of Goldstone's theorem in QCD, \emph{viz}.\\[-2ex]
\hspace*{2.5em}\parbox[t]{0.9\textwidth}{\flushleft \emph{Goldstone's theorem is fundamentally an expression of equivalence between the one-body problem and the two-body problem in QCD's colour-singlet pseudoscalar channel}.}

\medskip

Eqs.\,\eqref{gGTrelations} are a Class-A prediction.  They are model-independent, gauge-independent and scheme-independent: any continuum approach to bound-states in QCD that faithfully expresses and preserves chiral symmetry and the pattern by which it is broken will generate these identities.  There are numerous Class-B illustrations (see, e.g.\ Ref.\,\cite{Maris:1997tm} and references thereto).

It follows from Eqs.\,\eqref{gGTrelations} that pion properties are an almost direct measure of the mass function depicted in Fig.\,\ref{gluoncloud}.  Moreover, these identities are the reason the pion is massless in the chiral limit and indirectly the explanation for a proton mass of around 1\,GeV.  Thus, enigmatically, properties of the nearly-massless pion are the cleanest expression of the mechanism that is responsible for almost all the visible mass in the Universe.\footnote{Additional notable consequences of DCSB are described, \emph{e.g}.\  in Ref.\,\cite{Li:2016dzvFBS}, which explains some of
the results that follow from the fact that the leptonic decay constant of every radially-excited pseudoscalar meson must vanish in the chiral limit.}

Given its importance, one might ask whether the dressed-quark mass function is observable; and this issue is now readily addressed.  As just noted, Eqs.\,\eqref{gGTrelations} entail that in the neighbourhood of the chiral limit, the dressed-quark mass function (almost) completely determines the pion's Bethe-Salpeter wave-function.  Thus, like a wave function in any field of physics, the dressed-quark mass function is not strictly observable.  On the other hand, no one can reasonably doubt the enormous value of possessing (nearly) complete knowledge of a bound-state's wave function.
Moreover, the pion's Poincar\'e-covariant Bethe-Salpeter wave function can be projected onto the light-front.  The object thus obtained is strictly a probability amplitude and the moments of a probability measure are truly observable.  Consequently, there is a mathematically strict sense in which moments of the dressed-quark mass function are observable.  One should note in addition that generalised parton distributions can rigorously be defined as an overlap of light-front wave functions \cite{Burkardt:2000za, Diehl:2000xz, Burkardt:2002hr, Diehl:2003ny}.
Practically, therefore, the dressed-quark mass function can be ``measured'' because it influences and determines a vast array of experimental observables and there is at least one tractable framework, the DSEs, through which to relate those observables to QCD.  In this sense, $M(p^2)$ is as readily observable as, \emph{e.g}.\ the parton distribution amplitudes and functions which are the focus of a wide variety of extant and proposed experiments.

\subsection{Positive and negative contributions from the trace anomaly}

\begin{figure}[t]
\centerline{\includegraphics[width=0.60\textwidth]{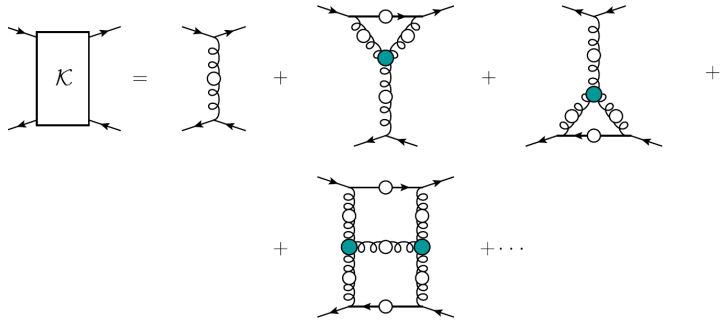}}
%
\caption{\label{FigEqGamma}
Some of the contributions to the quark-antiquark scattering kernel, $\mathpzc{K}$:
solid line with open circle, dressed-quark propagator; open-circle “spring”, dressed-gluon propagator, and shaded (blue) circle at the junction of three gluon lines, dressed three-gluon vertex.
The contribution drawn in the second row is an example of an H-diagram.  It is two-particle-irreducible and cannot be expressed as a correction to either vertex in a ladder kernel nor as a member of the class of crossed-box diagrams.%
(Additional details are provided in Ref.\,\cite{Binosi:2016rxz}.)}
\end{figure}

Return again now to Eq.\,\eqref{anomalypion}.  The pion's Poincar\'e-invariant mass and Poincar\'e-covariant wave function are obtained by solving a Bethe-Salpeter equation.  This is a scattering problem.  In the chiral limit, two massless fermions interact via exchange of massless gluons, \emph{i.e}.\ the initial system is massless; and it remains massless at every order in perturbation theory.  The complete calculation of the scattering process, however, involves an enumerable infinity of dressings and scatterings, as illustrated in Fig.\,\ref{FigEqGamma}.  This can be represented by a coupled set of gap- and Bethe-Salpeter equations.  At $\zeta=\zeta_2$, it is practical to build the kernels using a dressed-parton basis, \emph{viz}.\ from valence-quarks with a momentum-dependent running mass produced by self-interacting gluons, which have given themselves a running mass.

In the chiral limit one can prove algebraically \cite{Munczek:1994zz, Bender:1996bb, Chang:2009zb, Binosi:2016rxz} that, at any finite order in a symmetry-preserving construction of the kernels for the gap (quark dressing) and Bethe-Salpeter (bound-state) equations, there is a precise cancellation between the mass-generating effect of dressing the valence-quarks and the attraction introduced by the scattering events.  This cancellation guarantees that the simple system,  which began massless, becomes a complex system, with a nontrivial bound-state wave function that is attached to a pole in the scattering matrix, which remains at $P^2=0$, \emph{i.e}. the bound-state is also, therefore, massless.

The precise statement is that in the pseudoscalar channel, the dynamically generated mass of the two fermions is precisely cancelled by the attractive interactions between them, if and only if, Eqs.\,\eqref{gGTrelations} are satisfied.
It can be expressed as follows:
\begin{subequations}
\label{QuasiParticle}
\begin{align}
\langle \pi(q) | \theta_0 | \pi(q) \rangle
& \stackrel{\zeta \gg \zeta_2}{=} \langle \pi(q) | \tfrac{1}{4} \beta(\alpha(\zeta)) G^{a}_{\mu\nu}G^{a}_{\mu\nu} | \pi(q) \rangle  \stackrel{\zeta \simeq \zeta_2}{\to}
\langle \pi(q) | {\cal D}_1 + {\cal I}_2 |\pi(q)\rangle\,, \\
{\cal D}_1 & = \sum_{f=u,d} M_f(\zeta) \, \bar {\cal Q}_f(\zeta) {\cal Q}_f(\zeta) \,, \\
{\cal I}_2 &  = \tfrac{1}{4} [\beta(\alpha(\zeta)) {\cal G}^{a}_{\mu\nu}{\cal G}^{a}_{\mu\nu}]_{2{\rm PI}}  \,,
\end{align}
\end{subequations}
which describes the metamorphosis of the parton-basis chiral-limit expression of the expectation-value of the trace anomaly into a new expression, written in terms of a nonperturbatively-dressed quasi-particle basis, with dressed-quarks denoted by ${\cal Q}$ and the dressed-gluon field strength tensor by ${\cal G}$.
Here, the first term is positive, expressing the one-body-dressing content of the trace anomaly.  Plainly, a massless valence-quark (antiquark) acquiring a large mass through interactions with its own gluon field is an expression of the trace-anomaly in what might be termed the one-quasiparticle subsector of a complete pion wave function.
The second term is negative, expressing the two-particle-irreducible (2PI) scattering-event content of the forward scattering process represented by this expectation-value of the scale-anomaly.  This term acquires a scale because the couplings, and the gluon- and quark-propagators in the 2PI processes have all acquired a mass-scale.
Away from the chiral limit, and in other channels, such as the proton, the cancellation is incomplete.  It is worth reiterating that these statements can be verified algebraically.\footnote{Notably, if dealing with exotic or hybrid bound-states possessing a valence-gluon degree-of-freedom, then the ${\cal D}_1$ term in Eq.\,\eqref{QuasiParticle} would also include an explicit and positive gauge-invariant contribution associated with the dynamical generation of a dressed-gluon mass.  Moreover, such cancellations are a typical feature of bound-state problems, apparent already in both rudimentary and sophisticated uses of quantum mechanics \cite{Carlson:2014vla}.}

The fact that a Poincar\'e-invariant analysis of the simultaneous impact of the trace anomaly in the pion and proton cannot yield a sum of terms that is each individually positive precludes a useful ``pie diagram'' breakdown of the distribution of mass within a hadron.

\section{Conclusion}
Explanations of the origin of a hadron's mass and its distribution within that state depend on the observer's preferred frame of reference and resolving scale.  At a scale typical of contemporary continuum- and lattice-QCD ($\zeta=\zeta_2:=2\,$GeV), the dynamical chiral symmetry breaking (DCSB) paradigm provides an excellent and intuitive method to explicate and understand the associated, emergent phenomena.  This perspective leads to numerous predictions that can be tested at contemporary hadron physics facilities, related \emph{e.g}.\ to hadron elastic and transition form factors, and a diverse array of parton distribution amplitudes and functions.

At this point it is reasonable to enquire after the possible impact of new facilities, such as an electron ion collider (EIC), on our understanding of the emergence of mass in the Standard Model.  In response, one should first note that an EIC will enable access to a wide array of phenomena within the valence-quark domain, and, consequently, the predictions made for contemporary facilities are equally relevant for an EIC and might there be better be addressed.  In addition, hadron tomography will be a major focus of an EIC; and the DCSB paradigm has a great deal to say about that subject, providing a means for the QCD-connected computation of the pointwise behaviour of generalised and transverse-momentum-dependent parton distributions.  Finally, a major focus of an EIC will be low-$x$.  Gluons dominate in this region; but they also acquire a dynamically-generated mass.  That is likely to have a significant impact, \emph{e.g}.\ in connection with gluon saturation phenomena.  The challenge for theory now, therefore, is to identify those observables which are most sensitive to the emergence of mass as described above.  There is time; but also some urgency.

\begin{acknowledgements}
I would like to thank Z.-E.\,Meziani and J.-W.\,Qiu for organising the Temple Workshop: \emph{The Proton Mass: At the heart of most visible matter}, and they and the participants for hours of engaging discussions, which served to refocus my attention on some of the issues discussed herein.
Constructive comments were subsequently received from I.\,C.~Clo\"et, R.\,J.~Holt, V.~Mokeev, J.~Papavassiliou and J.~Rodr{\'i}guez-Quintero, and are also gratefully acknowledged.
Work supported by:
the U.S.\ Department of Energy, Office of Science, Office of Nuclear Physics, under contract no.~DE-AC02-06CH11357
\end{acknowledgements}



\end{document}